\documentclass[prb,aps,twocolumn,showpacs, nobibnotes,epsf]{revtex4}

\usepackage{graphicx}
\usepackage{dcolumn}
\usepackage{bm}
\usepackage{SIunits}
\usepackage{tabularx}

\begin{document}

\title{Phase diagram and physical properties of NaFe$_{1-x}$Cu$_x$As single crystals}

\author{A. F. Wang, J. J. Lin, P. Cheng, G. J. Ye, F. Chen, J. Q. Ma, X. F. Lu, B. Lei, X. G. Luo and X. H. Chen}

\altaffiliation{Corresponding author} \email{chenxh@ustc.edu.cn}

\affiliation{Hefei National Laboratory for Physical Science at Microscale and Department of
Physics, University of Science and Technology of China, Hefei, Anhui 230026, People's Republic
of China\\}

\begin{abstract}

A series of high quality NaFe$_{1-x}$Cu$_x$As single crystals has
been grown by a self-flux technique, which were systematically
characterized via structural, transport, thermodynamic, and high
pressure measurements. Both the structural and magnetic transitions
are suppressed by Cu doping, and bulk superconductivity is induced
by Cu doping. Superconducting transition temperature ($T_c$) is
initially enhanced from 9.6 to 11.5 K by Cu doping, and then
suppressed with further doping. A phase diagram similar to
NaFe$_{1-x}$Co$_x$As is obtained except that insulating instead of
metallic behavior is observed in extremely overdoped samples.
$T_c$'s of underdoped, optimally doped, and overdoped samples are
all notably enhanced by applying pressure. Although a universal
maximum transition temperature ($T_c^{max}$) of about 31 K under
external pressure is observed in underdoped and optimally doped
NaFe$_{1-x}$Co$_x$As, $T_c^{max}$ of NaFe$_{1-x}$Cu$_x$As is
monotonously suppressed by Cu doping, suggesting that impurity
potential of Cu is stronger than Co in NaFeAs. The comparison
between Cu and Co doping effect in NaFeAs indicates that Cu serves
as an effective electron dopant with strong impurity potential, but
part of the doped electrons are localized and do not fill the energy
bands as predicted by the rigid-band model.

\end{abstract}


\vskip 300 pt

\maketitle

\section{INTRODUCTION}

The discovery of iron-based superconductors gives another
opportunity to study the physics of high-temperature
superconductivity besides the cuprates.\cite{Hosono,Chen,BaK1} The
parent compounds of iron-based superconductors are antiferromagnetic
semimetals, and superconductivity can be induced by hole, electron
doping or applying pressure. For example, superconductivity was
induced in BaFe$_2$As$_2$ by the doping of Co, Ni, and
Cu.\cite{BaCo, Ni} In the case of Co and Ni doping, the doped
electron numbers predicted by rigid band model are $x$ and 2$x$,
respectively. The study on phase diagram of
Ba(Fe$_{1-x}$Co$_x$)$_2$As$_2$ and Ba(Fe$_{1-x}$Ni$_x$)$_2$As$_2$ by
angle-resolved photoemission spectroscopy (ARPES) indicates that the
doped electron number roughly follows the rigid-band
model.\cite{Fujimori} Therefore, it is natural to expect that
superconductivity could be induced by Cu doping and the doped
electron number is 3$x$. However, superconductivity was observed in
a very narrow range of doping in
Ba(Fe$_{1-x}$Cu$_x$)$_2$As$_2$,\cite{Ni} and no superconductivity
was observed in Sr(Fe$_{1-x}$Cu$_x$)$_2$As$_2$.\cite{Yan YJ}

Although the doping effect of Co and Ni is clear now, the role of Cu
doping is still under debate. X-ray photoelectron spectroscopy (XPS)
and x-ray absorption (XAS) measurements show that Cu 3$d$ states
locate at the bottom of the valence band in a localized 3$d^{10}$
shell, so that the formal valence state of Cu is $+1$ and the
substitution of Fe$^{2+}$ by Cu$^{1+}$ results in hole
doping.\cite{Yan YJ, Merz, JPCM} The theoretical and experimental
studies on SrCu$_2$As$_2$, the end member of
Sr(Fe$_{1-x}$Cu$_x$)$_2$As$_2$ series, indicate that it is an
$sp$-band metal with hole-type carries dominate and Cu in the
nonmagnetic 3$d^{10}$ electronic configuration corresponds to the
valence state Cu$^{1+}$,\cite{Yan YJ, Singh, Johnston} which further
supports the result of XPS and XAS. In addition, the doping effect
of Cu is similar to that of Mn, which is also considered as hole
doping and no superconductivity has been found.\cite{Canfield Mn} On
the other hand, electron doping by Cu was proved by
ARPES,\cite{Fujimori} and further confirmed by hall measurement on
Ba$_{0.6}$K$_{0.4}$(Fe$_{1-x}$Cu$_x$)$_2$As$_2$.\cite{Wen HH BaKCu}
Similar superconducting dome to that of
Ba(Fe$_{1-x}$Co$_x$)$_2$As$_2$ was observed in the phase diagram of
Ba(Fe$_{1-x-y}$Co$_x$Cu$_y$)$_2$As$_2$ (x $\sim$ 0.022 and x $\sim$
0.047),\cite{Ni} indicating the similar doping effect between Co and
Cu, and suggesting electron doping induced by Cu substitution.  To
reconcile this controversial situation, it is of great interest to
further investigate the Cu doping effect in other family of
iron-pnictide superconductors.

Besides BaFe$_2$As$_2$, high quality of NaFeAs single crystal is
also available now, which turns out to be suitable for studying the
Cu doping effect. NaFeAs is regarded as a filamentary
superconductor, and the superconductivity cannot be detected by
specific heat. By substituting Co or Ni on Fe sites, bulk
superconductivity was obtained.\cite{Parker, Wang AF Na} The Ni
doping doubles the amount of electron doping of Co,\cite{Parker}
which follows the rigid band model. Moreover, it has been reported
that $T_c$ of analogous compound LiFe$_{1-x}$Cu$_x$As is suppressed
linearly by Cu doping.\cite{LiFeCuAs} Hence, we study the role of Cu
doping in NaFe$_{1-x}$Cu$_x$As, and compare it with the effect of Cu
doping in BaFe$_2$As$_2$ and LiFeAs. In this paper, we report the
study on the physical properties and phase diagram of
NaFe$_{1-x}$Cu$_x$As by measuring x-ray diffraction (XRD),
resistivity, magnetic susceptibility, Hall coefficient, specific
heat, and high pressure. A phase diagram similar to
NaFe$_{1-x}$Co$_x$As is established. The comparison of the physical
properties and phase diagrams between NaFe$_{1-x}$Cu$_x$As and
NaFe$_{1-x}$Co$_x$As clearly indicates that Cu doping is electron
doping and the electron concentration deviates from the expected
3$x$. The deviation can be explained that part of the doped
electrons fill the impurity band which is located deep below the
Fermi level ($E_F$) and do not fill the energy bands as predicted by
the rigid-band model.\cite{Fujimori, JPCM}

\section{EXPERIMENTAL DETAILS}

A series of NaFe$_{1-x}$Cu$_x$As single crystals was grown by
adopting the NaAs flux method. Its growth procedure resembles to
that of NaFe$_{1-x}$Co$_x$As, and the details can be found in our
previous work.\cite{Wang AF Na} XRD was performed on a Smartlab-9
diffracmeter (Rikagu) from 10$^{\rm o}$ to 70$^{\rm o}$, with a
scanning rate of 6$^{\rm o}$ per minute. The actual chemical
composition of the single crystal was determined by energy
dispersive x-ray spectroscopy (EDX). The Cu content $x$ hereafter is
the actual composition determined by EDX. The resistivity and
specific heat measurements were carried out by using the PPMS-9T
(Quantum Design), and resistivity down to 50 mK were measured in a
dilution refrigerator on PPMS. The magnetic susceptibility was
measured using a vibrating sample magnetometer (VSM) (Quantum
Design). The Hall coefficient was measured on PPMS with the
four-terminal ac technique by switching the polarity of the magnetic
field $H$ //$c$ to remove any magnetoresistive components due to the
misalignment of the voltage contacts.\cite{Wang AF hall} The
pressure was generated in a Teflon cup filled with Daphne Oil 7373,
which was inserted into a Be-Cu pressure cell, and the pressure
applied in the resistivity measurement was determined by shifting
the superconducting transition temperature of pure Sn.

\section{RESULTS}

\begin{figure}[ht]
\centering
\includegraphics[width=0.49\textwidth]{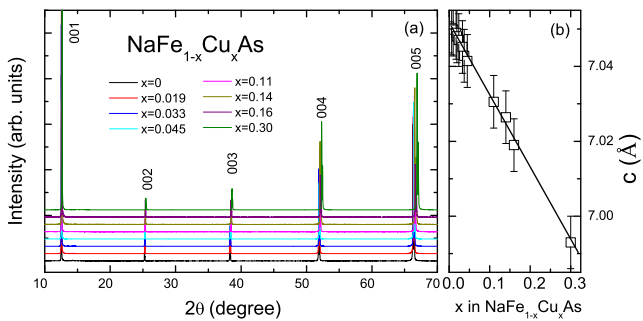}
\caption{(color online). (a) Selected XRD patterns for the
NaFe$_{1-x}$Cu$_x$As single crystals. (b) Doping dependence of the
c-axis parameter.} \label{fig1}
\end{figure}

\subsection{X-ray diffraction}

Figure 1(a) shows the selected single-crystalline XRD patterns for
the NaFe$_{1-x}$Cu$_x$As single crystals. Only (00\emph{l})
reflections can be recognized, indicating that the crystals are well
orientated along the \emph{c} axis. The lattice parameter $c$ is
estimated from the (00\emph{l}), and the evolution of all the single
crystals' lattice parameter $c$ with the doping level is shown in
Fig. 1(b). The lattice parameter $c$ decreases with increasing
doping concentration, which roughly obeys the Vegard's law.
Comparing with undoped NaFeAs, the amplitude of lattice parameter
change is about 0.8\% with Cu doping concentration up to 0.30, a
little smaller than that of Co doped NaFeAs with the same doping
level.\cite{Wang AF Na, Parker}

\begin{figure}[ht]
\centering
\includegraphics[width=0.49\textwidth]{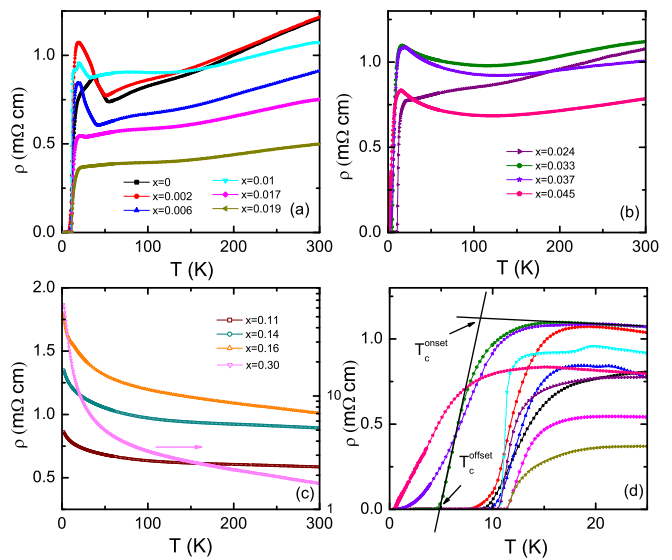}
\caption{(color online). (a)-(c) Temperature dependence of in-plane
resistivity for NaFe$_{1-x}$Cu$_x$As single crystals. (d)
Enlargement of the low temperature resistivity in panels (a) and
(b). The criteria used to determine the onset and offset temperature
for the superconducting transitions is also shown in (d).}
\label{fig2}
\end{figure}

\subsection{Electrical resistivity}

The temperature dependence of in-plane electrical resistivity for
NaFe$_{1-x}$Cu$_x$As single crystals are shown in Fig. 2. To make
the graphs easier to read, the data are grouped into three sets. The
resistivity at room temperature is about 0.4-1.7 m$\Omega$
cm.\cite{Wang AF Na, Spyrison} The error bar of the absolute
resistivity is relatively large comparing to the evolution of
resistivity caused by the doping effect, which is mainly coming from
the uncertainty of geometric factor. So we cannot observe a
systematic evolution of the room temperature resistivity in the
whole doping range. The superconducting transitions for most of the
samples are quite broad and the onset is very round, so we define
$T_c^{offset}$ as $T_c$, as shown in Fig. 2(d). $T_c$ stands for
$T_c^{offset}$ for convenience hereafter. The kinks associated with
the structural/spin density wave (SDW) transition are clearly
resolved in the low temperature resistivity of underdoped
crystals.\cite{Dai PC} We use the same criteria to define the
structural and SDW transition as described in Ref. 15.

The structural and SDW transitions are progressively suppressed with
increasing Cu concentration, similar to NaFe$_{1-x}$Co$_x$As. $T_c$
increases slightly with Cu doping in the underdoped region. The
maximum $T_c$ about 11.5 K is reached at $x$ = 0.019, but the
amplitude of $T_c$ enhancement (2K) is much smaller than that in Co
doped NaFeAs (10K). $T_c$ decreases quickly with further increasing
Cu doping, and no trace of superconducting transition is observed in
crystals with doping concentration larger than 0.045.
Metal-insulator transition is observed in the crystals with doping
level higher than 0.033. Ultimately, insulating behavior in the
whole temperature range is observed in the extremely overdoped
crystals, which is quite different from the metallic behavior in
extremely overdoped NaFe$_{1-x}$Co$_x$As.\cite{Wang AF Na} A weak
semiconducting behavior is also observed in
Ba(Fe$_{1-x}$Cu$_x$)$_2$As$_2$ and
Sr(Fe$_{1-x}$Cu$_x$)$_2$As$_2$.\cite{Ni, Yan YJ} In addition, when
$\sim$ 4\% Fe was substituted by Cu, a metal-insulator transition
was observed in Fe$_{1.01-x}$Cu$_x$Se.\cite{Williams, Huang} It is
reported that the insulator phase of Fe$_{1.01-x}$Cu$_x$Se is an
Anderson localized system arising from disorder rather than a
conventional semiconductor.\cite{Chadov} Whether the metal-insulator
transition in NaFe$_{1-x}$Cu$_x$As and Fe$_{1.01-x}$Cu$_x$Se have
common origination need further investigation.

\begin{figure}[ht]
\centering
\includegraphics[width=0.49\textwidth]{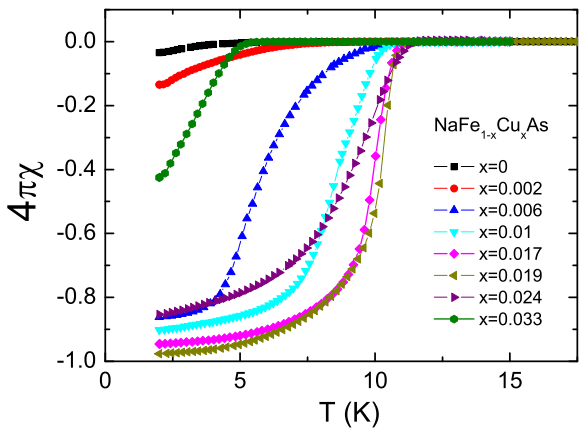}
\caption{(color online). Temperature dependent magnetic susceptibility for NaFe$_{1-x}$Cu$_x$As with magnetic field 10 Oe} \label{fig3}
\end{figure}

\subsection{Magnetic susceptibility}

Figure 3 shows the zero-field-cooling (ZFC) magnetic susceptibility
taken at 10 Oe with $H$ perpendicular to the $c$ axis for the
superconducting NaFe$_{1-x}$Cu$_x$As single crystals. As reported
previously, a tiny diamagnetic signal was observed below 9 K in
undoped NaFeAs.\cite{Wang AF Na} With Cu doping, the superconducting
shielding fraction rises rapidly. Bulk superconductivity with large
shielding fraction is observed in the composition range of 0.006
$\sim$ 0.024, indicating that Cu doping is beneficial to the
superconductivity of NaFeAs. $T_c$ inferred from the diamagnetic
signal is consistent with that determined by resistivity
measurement.  As shown in Fig. 3, $x$ $=$ 0.019 is the optimally
doped composition with maximum $T_c$ and largest shielding fraction.
Both shielding faction and $T_c$ decrease with further Cu doping,
and no diamagnetic signal above 2 K is detected in samples with the
Cu doping level higher than 0.033.

\begin{figure}[ht]
\centering
\includegraphics[width=0.49\textwidth]{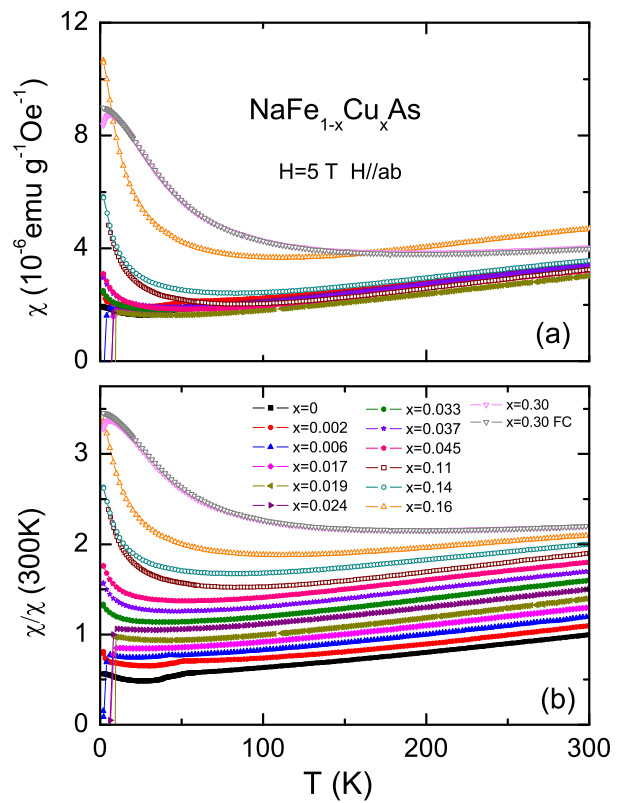}
\caption{(color online). (a) Temperature dependence of normal state magnetic
susceptibility for NaFe$_{1-x}$Cu$_x$As under a magnetic field of 5 T.
(b) Temperature dependence of the normalized magnetic susceptibility,
which is shifted upward by 0.1 for clarity.} \label{fig4}
\end{figure}

Figure 4 (a) presents the normal state in-plane magnetic
susceptibility for NaFe$_{1-x}$Cu$_x$As under a magnetic field of 5
T. The magnitude and behavior is similar to that in Co doped
NaFeAs.\cite{Wang AF Na} Because the magnitude of $\chi$ does not
change much in the doping range up to 0.014 (about 15\%), the
normalized susceptibility is shown in Fig. 4(b) for clarity.  All
the magnetic susceptibility is taken under zero-field-cooled (ZFC)
mode, and a field-cooled (FC) susceptibility of
NaFe$_{0.70}$Cu$_{0.30}$As is also presented. Rapid drops associated
with superconducting transition can still be observed at low
temperature for the superconducting samples. Kinks corresponding to
the structural and SDW transitions are observed in the undoped and
underdoped samples, which are suppressed with Cu doping and
consistent with the observation of resistivity. It is worth noting
that $\chi$ shows an almost linear temperature dependence in high
temperature for concentration up to 0.16. The slope of the linear
dependence of high-temperature susceptibility slightly decreases
with Cu doping, similar to Co doping.\cite{Wang AF Na} The linear
temperature dependence of high temperature susceptibility is a
common feature in iron-based superconductors, which has been
observed in NaFe$_{1-x}$Co$_x$As,\cite{Wang AF Na}
Ba(Fe$_{1-x}$Co$_x$)$_2$As$_2$,\cite{Wang XF Co} and
LaFeAsO$_{1-x}$F$_x$.\cite{LaOFeAs M} An explanation based on the
$J_1$-$J_2$ model of localized spins ascribes this behavior to the
spin fluctuations arising from the local SDW correlation.\cite{Zhang
GM} It is also argued that the behavior can be explained based on
the spin susceptibiltiy of a 2D Fermi-liquid with nearly nested
electron and hole pockets of the Fermi surface.\cite{Korshunov}
There is a Curie-Weiss like upturn in the low temperature magnetic
susceptibility of overdoped NaFe$_{1-x}$Cu$_x$As, which has been
reported in many Fe-based superconductors.\cite{Wang AF Na, Ni,
Xiang ZJ, Cao GH} The susceptibility upturn is usually attributed to
extrinsic origin, such as defects or impurities. A small separation
between the ZFC and FC occurs at NaFe$_{0.70}$Cu$_{0.30}$As, which
is considered as spin glass transition.\cite{Yan YJ}

\begin{figure}[ht]
\centering
\includegraphics[width=0.49\textwidth]{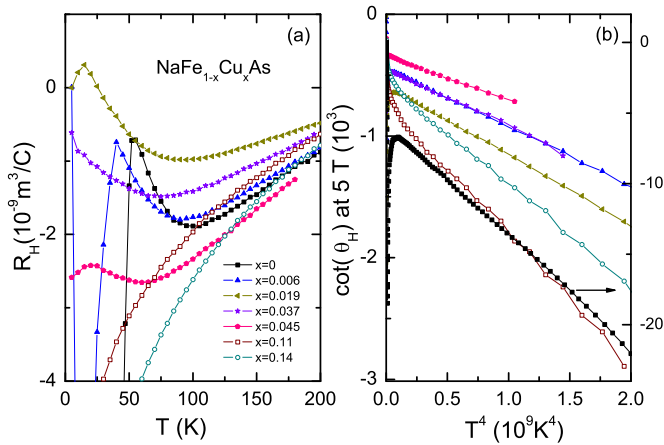}
\caption{(color online). (a) Temperature dependence of the Hall
coefficient $R_H$ for selected NaFe$_{1-x}$Cu$_x$As single crystals.
(b) The cot$\theta_H$ of NaFe$_{1-x}$Cu$_x$As single crystals plotted
in power-law temperature scale. The legends in (b) are same as (a)} \label{fig5}
\end{figure}

\subsection{Hall effect}

Figure 5(a) shows the temperature dependence of Hall coefficients
for selected single crystals of NaFe$_{1-x}$Cu$_x$As. The sharp drop
in the $x$ = 0 and $x$ = 0.006 single crystals is due to structural
transition, and the temperature of the transition determined from
resistivity, magnetic susceptibility, and Hall coefficient is
consistent with each other. The negative Hall coefficient of all the
single crystals indicate that the dominated carrier is electron. A
systematic evolution is observed on the absolute value of Hall
coefficient at 200 K. The value decreases with Cu doping up to
optimally doped crystal with x=0.019, and then increases with
further Cu doping. Due to the multiband effect and different
mobility of electron and hole carries, the Hall behavior is complex.
If we simply take the single band expression $n_H$ =
1/($eR_H$),\cite{Rullier Albenque} the behavior of Hall coefficient
of NaFe$_{1-x}$Cu$_x$As indicates that the Cu doping is electron
doping, similar to that of Co doping in
NaFe$_{1-x}$Co$_x$As.\cite{Wang AF Na} As the Cu doping, Hall
coefficients of Ba$_{0.6}$K$_{0.4}$(Fe$_{1-x}$Cu$_x$)$_2$As$_2$
gradually change from positive values to negative values, clearly
showing electron carriers are introduced. The result of Hall
measurement on NaFe$_{1-x}$Cu$_x$As is similar to the result of
Ba$_{0.6}$K$_{0.4}$(Fe$_{1-x}$Cu$_x$)$_2$As$_2$.\cite{Wen HH BaKCu}

As shown in Fig. 5(b), The Hall angle is plotted as cot$\theta_H$ =
$\rho$/$\rho_{xy}$ vs $T^4$, where $\rho$ is in-plane resistivity
and $\rho_{xy}$ is Hall resistivity. It has been reported that
cot$\theta_H$ shows power-law temperature dependence for all the
single crystals of NaFe$_{1-x}$Co$_x$As: $T^4$ for the parent
compound, approximately $T^3$ for the superconducting crystals and
$T^2$ for the heavily overdoped non-superconducting
sample.\cite{Wang AF hall} But $T^{\beta}$-dependent cot$\theta_H$
with $\beta$ $\approx$ 4 is observed in these NaFe$_{1-x}$Cu$_x$As
single crystals. This value is different from NaFe$_{1-x}$Co$_x$As,
Ba(Fe$_{1-x}$Co$_x$)$_2$As$_2$,\cite{JSNM} and hole doped
cuprates,\cite{Chien} but similar to electron-doped
cuprates.\cite{Wang CH} Since the $T^4$-dependence in cuprate is
interpreted by the multi-band effect with different contributions
from various bands, the different power law dependence between
NaFe$_{1-x}$Cu$_x$As and NaFe$_{1-x}$Co$_x$As indicates the
different band evolution of NaFeAs by Cu/Co doping.

\begin{figure}[ht]
\centering
\includegraphics[width=0.49\textwidth]{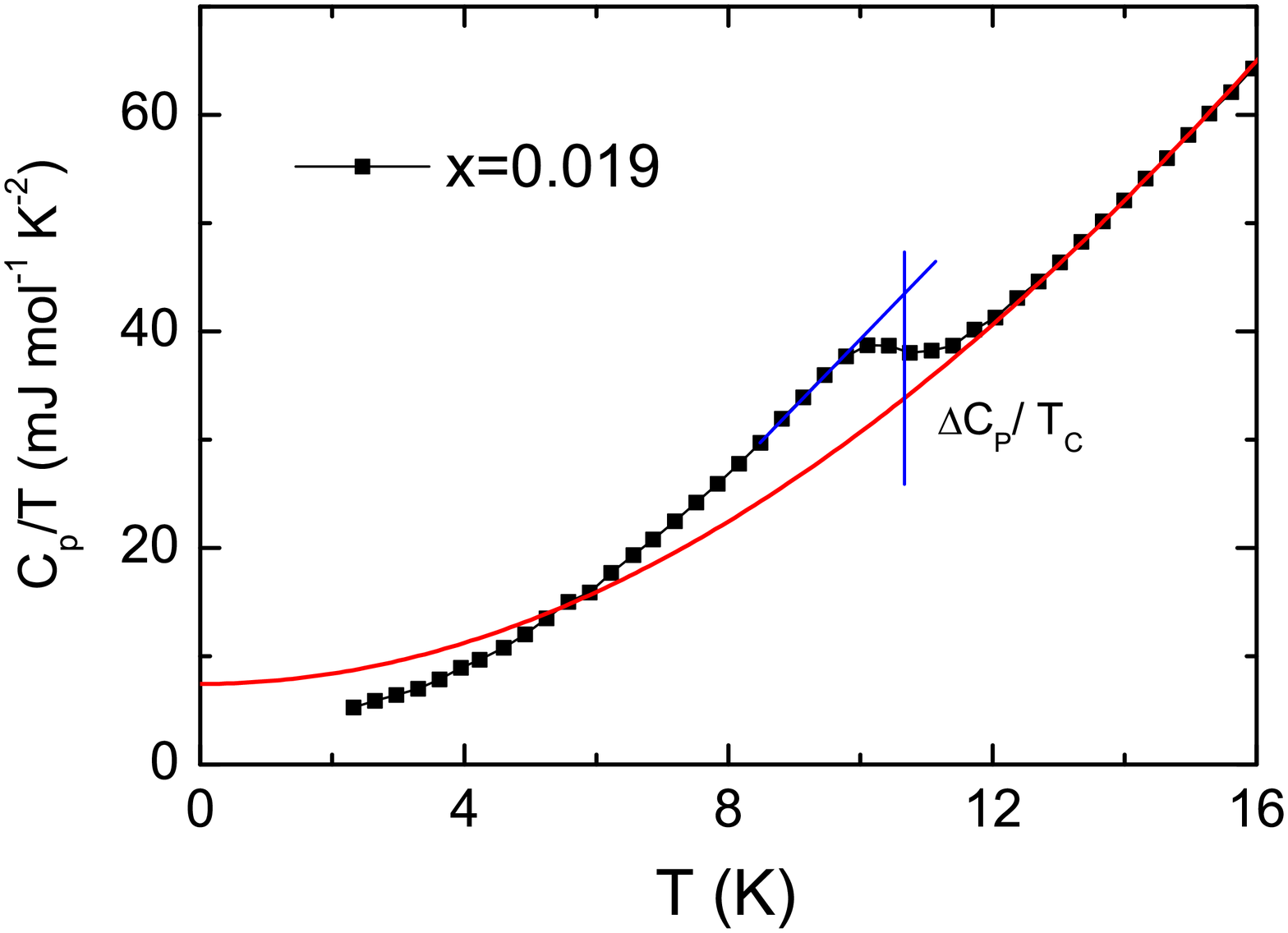}
\caption{(color online). Specific heat of optimally doped NaFe$_{0.981}$Cu$_{0.019}$As.
Blue lines show how the $\Delta$$C_p/T_c$ was determined. The red line is the best fit
of the specific heat between 13 and 30 K. (Only data below 16 K is shown for clarity )} \label{fig6}
\end{figure}

\subsection{Specific heat}

To verify the bulk thermodynamic nature of the superconducting
transition, specific heat measurement were performed on optimally
doped NaFe$_{0.981}$Cu$_{0.019}$As, as shown in Fig. 6. Remarkable
jump in specific heat corresponding to superconducting transition is
observed, while no anomaly can be observed on NaFeAs single crystal.
The obvious jump in specific heat suggests that bulk
superconductivity in NaFeAs is obtained by Cu doping.

The red line is the best fit of the normal state specific heat
between 13 and 30 K by $C_P$ = $\gamma_n$$T$ + $\beta$$T^3$ +
$\eta$$T^5$, where $\gamma_n$$T$ and $\beta$$T^3$ + $\eta$$T^5$ are
electron and phonon contributions, respectively. It is found that
$\gamma_n$ = 7.44 mJ mol$^{-1}$ K$^{-2}$, $\beta$ = 0.238 mJ
mol$^{-1}$ K$^{-4}$, and $\eta$ = -5.00 $\times$ 10$^{-5}$ mJ
mol$^{-1}$ K$^{-6}$. The estimated Debye temperature is 290 K,
almost same to the corresponding value for
NaFe$_{0.972}$Co$_{0.028}$As.\cite{Wang AF Na}  $\Delta$$C_p/T_c$ at
$T_c$ = 10.65 K is estimated to 9.85 mJ mol$^{-1}$ K$^{-2}$ by
isentropic construction sketched in Fig. 6. The value roughly follow
the expanded BNC scaling, which is proposed by Bud'ko, Ni, and
Canfield (BNC) and expanded by J. S. Kim $\emph{et al.}$\cite{BNC1,
BNC2} The BNC scaling is considered as a simple text of whether a
material belongs to the iron-based superconductors.\cite{Stewart} So
the pairing symmetry of NaFe$_{0.981}$Cu$_{0.019}$As may be similar
to Ba(Fe$_{0.925}$Co$_{0.075}$)$_2$As$_2$ and
NaFe$_{0.972}$Co$_{0.028}$As. It has been reported that
$\Delta$$C_p$/$\gamma_n$$T_c$ = 2.11 in
NaFe$_{0.972}$Co$_{0.028}$As.\cite{Wang AF Na}  Base on the data
obtained above, $\Delta$$C_p$/$\gamma_n$$T_c$ in
NaFe$_{0.981}$Cu$_{0.019}$As is estimated to 1.32, a little smaller
than 1.43 expected for weak-coupling BCS superconductor.  The
different value between optimally Cu and Co doped NaFeAs suggests
that the coupling strength in Cu doped NaFeAs is weaker than Co
doped. As a result, NaFe$_{0.981}$Cu$_{0.019}$As may be a two band
$s$-wave superconductor with a weaker coupling strength.

  \begin{figure}[ht]
  \centering
  \includegraphics[width=0.49\textwidth]{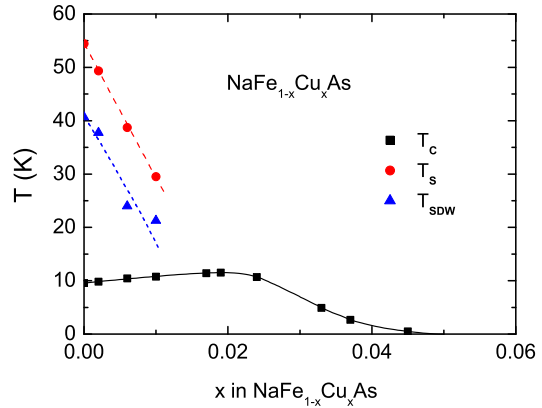}
  \caption{(color online). $T-x$ phase diagram of NaFe$_{1-x}$Cu$_x$As. Lines are guides to the eye.} \label{fig7}
  \end{figure}

\subsection{Phase diagram}

The $T-x$ phase diagram of NaFe$_{1-x}$Cu$_x$As is plotted in Fig.
7, where $T_s$, $T_{SDW}$, and $T_c$ stand for structural, SDW, and
superconducting transition, respectively. The data in Fig. 7 is
obtained from resistivity, which is consistent with magnetic
susceptibility, Hall, and specific heat measurements. As the Cu
doping, both structural and SDW transitions are progressively
suppressed to low temperature, and $T_c$ is enhanced slightly from
9.6 K in NaFeAs to 11.5 K in optimally Cu doped NaFeAs. $T_c$
decreases with Cu concentration in overdoped region, and
metal-insulator transition is observed in the normal state
resistivity of overdoped superconducting samples. Extremely
overdoped samples exhibit insulating behavior, which is different
from the overdoped nonsuperconducting NaFe$_{1-x}$Co$_x$As, where
metallic behavior is observed. A weak semiconductor behavior is also
observed on overdoped Ba(Fe$_{1-x}$Cu$_x$)$_2$As$_2$ and
Sr(Fe$_{1-x}$Cu$_x$)$_2$As$_2$.\cite{Ni, Yan YJ} Although the
magnitude of $T_c$ enhancement of NaFeAs is only 1.9 K by Cu doping,
the negligible small shielding fraction is greatly enhanced to
nearly 100\%. The full shielding fraction indicates Cu doping is
beneficial for the superconductivity of NaFeAs, contrast to the case
of LiFe$_{1-x}$Cu$_x$As.\cite{LiFeCuAs} But the maximum $T_c$ is
obviously lower than 20 K in NaFe$_{1-x}$Co$_x$As under ambient
pressure, the maxiumum $T_c$ may be suppressed by stronger impurity
potential of Cu. Microscopic coexistence of SDW and
superconductivity has been proved by scanning tunneling microscopy
(STM) in underdoped NaFe$_{1-x}$Co$_x$As.\cite{Wang YY} As shown in
Fig. 7, superconductivity also coexists with SDW in underdoped
NaFe$_{1-x}$Cu$_x$As.

  \begin{figure}[ht]
  \centering
  \includegraphics[width=0.49\textwidth]{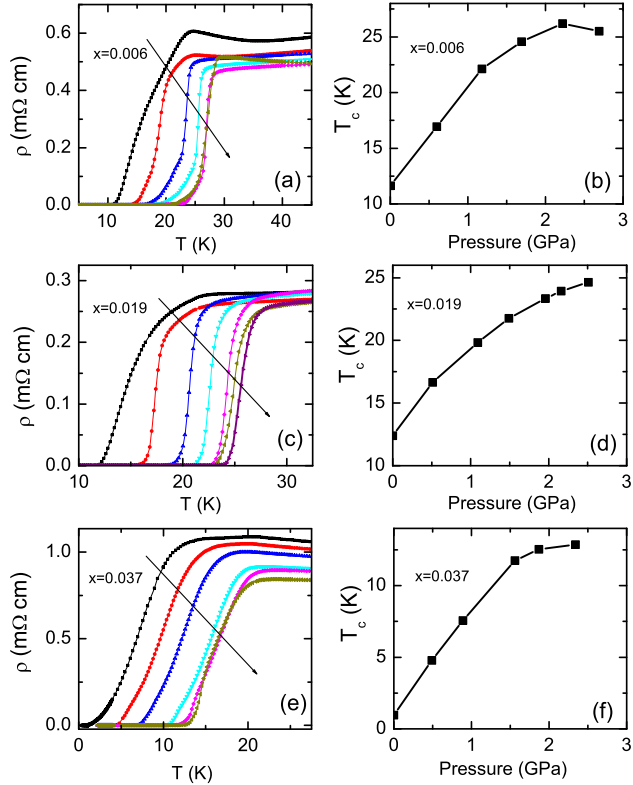}
  \caption{(color online). Left panels: in-plane resistivity of NaFe$_{1-x}$Cu$_x$As
  (x=0.006 (a), 0.019 (c) and 0.037 (e)) under various pressures, arrows indicate the
  direction of the increasing pressure. Right panels: the $T(p)$ phase diagrams of the
  samples corresponding to the left panels.} \label{fig8}
  \end{figure}

  \begin{figure}[ht]
  \centering
  \includegraphics[width=0.49\textwidth]{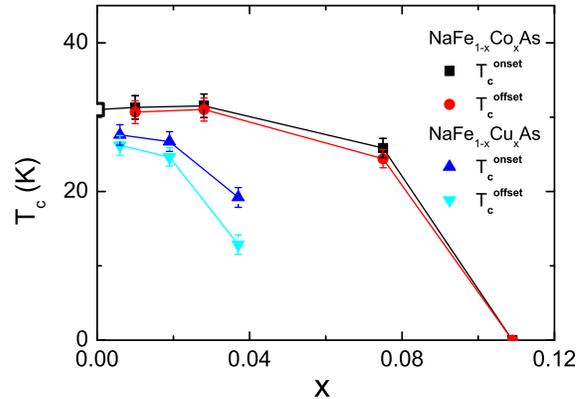}
  \caption{(color online). $T_c^{max}$ of NaFe$_{1-x}$Co$_x$As and NaFe$_{1-x}$Cu$_x$As
  plotted as functions of the Co/Cu substitution, $x$. The open black symbol represents
  the $T_c^{max}$ of NaFeAs reported by Zhang $\emph{et al}$.\cite{Jin CQ} The data
  of NaFe$_{1-x}$Co$_x$As is taken from Ref. 37.}
  \label{fig9}
  \end{figure}

  \subsection{Pressure effects}

Resistivity measurements under pressure were performed in underdoped
sample with $x$ = 0.006, optimally doped $x$ = 0.019, and overdoped
$x$ = 0.037 samples. As shown in Fig. 8 (a), for underdoped single
crystal with $x$ = 0.006, the resistivity upturn associated with the
structural or SDW transitions is suppressed to low temperature by
pressure and eventually become indistinguishable. $T_c$ initially
increased by applying pressure, and maximum $T_c$ = 26.2 K is
observed at 2.2 GPa, the $T_c$ decreases with further increasing
pressure, the data is summarized in Fig. 8(b). For optimally doped
and overdoped samples, where structural and SDW transitions have
been suppressed by Cu doping, $T_c$ are monotonously enhanced up to
the maximum pressure in our measurement.

As shown in Figs. 8(d) and 8(f), the $T_c^{max}$ = 24.6 and 12.9 K
are obtained for $x$ = 0.019 and $x$ = 0.037, respectively. The
pressure coefficient between ambient pressure and the pressure at
which $T_c$ reaches its maximum is 5.2, 4.9, and 5.1 K GPa$^{-1}$,
comparable to the optimally doped and overdoped
NaFe$_{1-x}$Co$_x$As.\cite{Wang AF pressure}

The maximum transition temperature ($T_c^{max}$) of
NaFe$_{1-x}$Co$_x$As and NaFe$_{1-x}$Cu$_x$As obtained under
pressure is plotted on Fig. 9. As shown in Fig. 9, the maximum $T_c$
obtained by combining the effect of doping and pressure in
NaFe$_{1-x}$Cu$_x$As decreases with Cu concentration, contrast to
the pressure effect on NaFe$_{1-x}$Co$_x$As, where maximum $T_c$
about 31 K is observed from undoped to optimally doped
samples.\cite{Wang AF pressure} Substitution of Fe by other
transition metal can induce carries as well as impurities. Because
both Co and Cu can dope electron into NaFeAs, it is mainly the
impurity effect that responds for the different $T_c^{max}$
evolution as a function of doping. If we only consider the impurity
effect on the maximum transition temperature of NaFeAs,
$T_c$-suppression rate for Cu is $\Delta$$T_c$/Cu-1\% = -4.3 K,
slightly larger than -3.5 K in
Ba$_{0.6}$K$_{0.4}$(Fe$_{1-x}$Cu$_x$)$_2$As$_2$,\cite{Wen HH BaKCu}
but much larger than -0.7 K in NaFe$_{1-x}$Co$_x$As with similar
doping concentration. The large $T_c$-suppression rate of
NaFe$_{1-x}$Cu$_x$As suggests that the impurity potential of Cu is
stronger than Co in NaFeAs.

\begin{figure}[ht]
\centering
\includegraphics[width=0.49\textwidth]{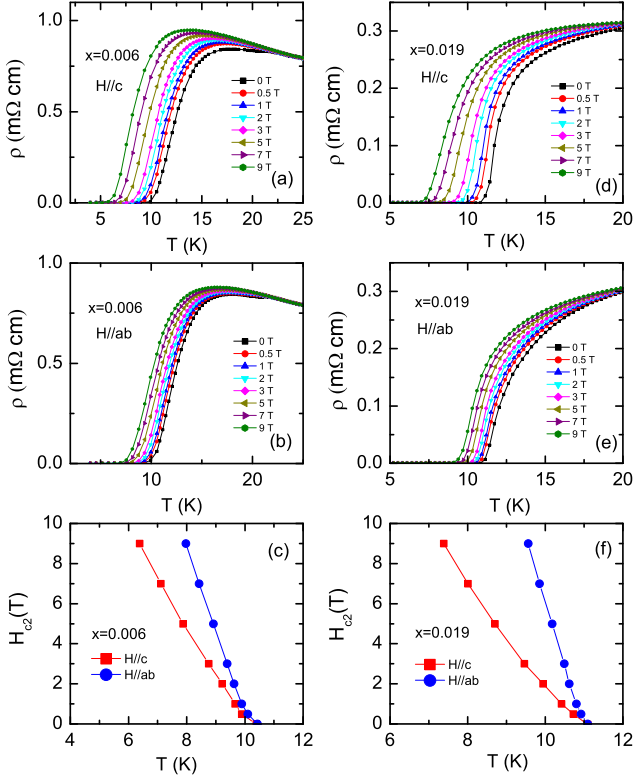}
\caption{(color online). The temperature dependence of resistivity for underdoped
NaFe$_{0.994}$Cu$_{0.006}$As (a)-(b), and optimally doped NaFe$_{0.981}$Cu$_{0.019}$As
(d)-(e) crystals with the magnetic field parallel and perpendicular to the $c$ axis, respectively.
(c) and (f) The temperature dependence of $H_{c2}$ for NaFe$_{0.994}$Cu$_{0.006}$As and NaFe$_{0.981}$Cu$_{0.019}$As, respectively.} \label{fig10}
\end{figure}

\subsection{Anisotropy of the upper critical field}

In Fig. 10, we present the temperature dependence of resistivity for
NaFe$_{1-x}$Cu$_x$As (x=0.006, 0.019) under various magnetic fields.
The transition temperature of superconductivity (the criteria is
shown in Fig. 2(d)) is suppressed gradually and the transition is
broadened with increasing magnetic field. The effect of magnetic
field is much larger when the field is applied along the $c$ axis of
the single crystals instead of within the $ab$ plane. For underdoped
sample NaFe$_{0.994}$Cu$_{0.006}$As, the positive magnetoresistance
appears well below the temperature which is defined as the SDW
transition. The similar phenomenon was observed in NaFeAs single
crystal,\cite{Chen GF} and confirmed by neutron scattering.\cite{Dai
PC} Temperature dependent $H_{c2}$ curves for
NaFe$_{0.994}$Cu$_{0.006}$As and NaFe$_{0.981}$Cu$_{0.019}$As is
shown in Figs. 7(c) and 7(f), respectively. In order to determine
the upper critical field in the low-temperature region, we adopt the
Werthamer-Helfand-Hohenberg (WHH) formula $H_{c2}$(0) =
0.693[-($dH_{c2}/dT$)]$_{T_c}$$T_c$ for single band BCS
superconductor. We obtain [-($dH_{c2}^{ab}$/$dT$)]$_{T_c}$ = 4.23
T/K and [-($dH_{c2}^{c}$/$dT$)]$_{T_c}$ = 2.24 T/K at $T_c$ = 10.40
K from Fig. 7(c) for NaFe$_{0.994}$Cu$_{0.006}$As, so the
$H_{c2}$(0) can be estimated to be 30 and 16 T with the field
parallel and perpendicular to the $ab$ plane, respectively. In the
same way, $H_{c2}^{ab}$(0) = 49 T and $H_{c2}^{c}$(0) = 22 T are
obtained for NaFe$_{0.981}$Cu$_{0.019}$As single crystal. As a
result, the anisotropy parameter $\gamma_H$ =
$H_{c2}^{ab}$(0)/$H_{c2}^{c}$(0) can be estimated to be 1.88 and
2.22 for NaFe$_{0.994}$Cu$_{0.006}$As and
NaFe$_{0.981}$Cu$_{0.019}$As, respectively. The smaller anisotropy
$\gamma_H$ of underdoped samples than the overdoped samples has also
observed on Ba(Fe$_{1-x}$Co$_x$)$_2$As$_2$, where $\sim$ 50\%
smaller anisotropy was found in underdoped region.\cite{Ni N Co}
These anisotropy values are close to 2.25 - 2.35 for
NaFe$_{1-x}$Co$_x$As,\cite{Spyrison} and a little larger than 1.7 -
1.86 in Ba$_{0.60}$K$_{0.40}$Fe$_2$As$_2$ and Fe(Se, Te)
system,\cite{Yuan HQ, Fang MH} but samaller than 5 - 9 in
NdFeAsO$_{1-x}$F$_x$.\cite{Wen HH}

  \section{Discussion}

Although the $T_c$ of optimally doped NaFe$_{1-x}$Cu$_x$As is lower
than that in NaFe$_{1-x}$Co$_x$As and the superconducting dome is much narrower than that of  NaFe$_{1-x}$Co$_x$As, the
overall phase diagram of NaFe$_{1-x}$Cu$_x$As is similar to those of
NaFe$_{1-x}$Co$_x$As and Ba(Fe$_{1-x-y}$Co$_x$Cu$_y$)$_2$As$_2$ ($x$
$\sim$ 0.022 and 0.047).\cite{Wang AF Na, Ni} This indicates that
the similar doping effect of Cu and Co. As a result, the Cu doping
definitely introduces electron carries into NaFe$_{1-x}$Cu$_x$As.
The main difference between Cu and Co doped NaFeAs lies in that the
insulating phase instead of the metallic phase is observed in the
extremely overdoped samples, indicating Cu doping effect distinct
from Co except carries doping effect.

According to the comparison of the phase diagrams for
Ba(Fe$_{1-x}$TM$_x$)$_2$As$_2$ (TM = Co, Ni, Cu, and Co/Cu), the
narrow superconducting dome of Ba(Fe$_{1-x}$Cu$_x$)$_2$As$_2$ has
been interpreted that too many electrons have been added when the
structural/antiferromagnetic phase transitions are suppressed low
enough.\cite{Ni} But recent ARPES result on
Ba(Fe$_{1-x}$TM$_x$)$_2$As$_2$ (TM = Co, Ni, and Cu) suggests that
although electrons are indeed doped, part of them may be localized
and do not fill the energy bands as predicted by the rigid-band
model.\cite{Fujimori} Theory calculation found that the substitution
with strong impurity potential induces an impurity band split-off
below the original host band, which reduces the electron occupy from
the host band, result in decrease of the electron
occupation.\cite{Wadati, Berlijn, Nakamura} ARPES and density
functional theory (DFT) studies found that the impurity potential of
the substituted atoms enhances from Co, Ni, to Cu.\cite{Fujimori,
Wadati, Goldman} As a result, the number of electron doped by Cu is
less than the value expected from the simple rigid-band model.

As also suggested by high pressure measurement, the impurity effect
of Cu is stronger than Co. The dome of superconductivity is mainly
controlled by the balance of carrier concentration and impurity
scattering induced by the dopants. So the narrow or even absence of
superconducting dome in the phase diagram of Cu doped BaFe$_2$As$_2$
arises from that enough carriers are doped and so many impurities
are induced. Therefore, the expanded superconducting dome of
Ba(Fe$_{1-x-y}$Co$_x$Cu$_y$)$_2$As$_2$ ($x$ $\sim$ 0.022 and 0.047)
than Ba(Fe$_{1-x}$Cu$_x$)$_2$As$_2$ can be understood that fewer
carriers are needed when BaFe$_2$As$_2$ have been electron doped
with Co. While in NaFeAs, whose structural/SDW transition
temperature is much lower than BaFe$_2$As$_2$, fewer electrons are
required to suppress SDW and induce superconductivity. Therefore, Cu
doping can provide enough carriers to map out a phase diagram
similar to NaFe$_{1-x}$Co$_x$As and Ba(Fe$_{1-x}$Co$_x$)$_2$As$_2$.
Meanwhile, as the Cu concentration further increases, density of
States (DOS) at $E_F$ is gradually removed by the impurity band
induced by Cu doping. As a result, metal-insulator transition is
observed as a function of doping.

This scenario can also explain the contradiction between results of
XPS/XAS and ARPES. Since Cu 3$d$ states are located deeper below the
$E_F$ than Co,\cite{Fujimori, JPCM, Wadati} the extra $d$ electrons
for Cu almost totally locate around the substituted site. Hence,
closed 3$d$ shell is observed by XPS and XAS, although there is a
little delocated electron introduced by Cu dopant. It is found that
substitutions of Cu for Fe in (Ba, Sr)Fe$_2$As$_2$ at low level
result in electron doping, while in SrCu$_2$As$_2$, the end member
of Sr(Fe$_{1-x}$Cu$_x$)$_2$As$_2$,  is an $sp$-band metal with
hole-type carries dominate.\cite{Yan YJ, Fujimori, Johnston} The
contradictory result has been interpreted that there is a crossover
between electron and hole doping with increasing $x$, which is
induced by tetragonal (T) to collapsed tetragonal (cT) phase
transition as a function of doping.\cite{Johnston} Thus, in the case
of NaFe$_{1-x}$Cu$_x$As, where no cT phase has been observed, it is
natural to observe electron doping at low-level substitution of Cu
for Fe.

\section{SUMMARY AND CONCLUSIONS}

In conclusion, we have performed structural, transport,
thermodynamic, and high pressure measurements on
NaFe$_{1-x}$Cu$_x$As single crystals. Enough carries can be provided
by Cu doping to map out a phase diagram similar to
NaFe$_{1-x}$Co$_x$As.  In underdoped region, Both the structural and
SDW transition are monotonically suppressed by Cu doping. $T_c$ and
the superconducting shielding fraction are enhanced with the doping.
Bulk superconductivity with $T_c$ = 11.5 K is observed at optimally
doped sample, and a metal-insulator transition is observed with
further doping. Finally, insulating instead of metallic behavior in
NaFe$_{1-x}$Co$_x$As is observed in extremely overdoped
non-superconducting samples. $T_c$ is obviously enhanced by
pressure, but the $T_c^{max}$ decreases with Cu concentration. The
Hall measurements and comparison between Cu and Co doped NaFeAs
phase diagrams indicate that Cu doping introduces electron into
system, but the number of electron is far from 3$x$ as predicted by
rigid-band model.

\section*{Acknowledgements}

  This work is supported by the National Natural Science Foundation of China
(Grants No. 11190021, 11174266, 51021091), the "Strategic Priority
Research Program (B)" of the Chinese Academy of Sciences (Grant No.
XDB04040100), the National Basic Research Program of China (973
Program, Grants No. 2012CB922002 and No. 2011CBA00101), and the
Chinese Academy of Sciences.

\end{document}